\documentclass[aps,twocolumn,superscriptaddress,amsmath,showpacs,tightenlines]{revtex4}
\usepackage{graphicx}
\usepackage{epsfig}
\begin{document}
%++++++++++++++++
\title{Numerical approach for retention characteristics of double floating-gate memories}

\author{Tetsufumi Tanamoto and Kouichi Muraoka}
\affiliation{Corporate R \& D center, Toshiba Corporation,
Saiwai-ku, Kawasaki 212-8582, Japan}

\date{\today}

\begin{abstract}
We report on a numerical investigation in which memory characteristics of double floating-gate (DFG) structure
were compared to those of the conventional single floating-gate structure, including an interference effect
between two cells. 
We found that the advantage of  the DFG structure is its longer retention time and the disadvantage 
is its smaller threshold voltage shift.
We also provide an analytical form of charging energy including the interference effect.
%By analyzing the charging energy of two-cell DFGs, we show that the longer retention time 
%is caused by the increase of the charging energy.
\end{abstract}
%\pacs{03.67.Lx, 03.67.Mn, 73.21.La}
\maketitle

%=========================================%
% \section{Introduction}
Floating-gate (FG) memories are widely used in computers 
because of their low cost and high density~\cite{noguchi,Kim}.
FG memories have progressed rapidly through down-scaling using 
state of the art technology. However, several problems have been arising as a result of the 
progress of down-scaling of the FG structure to the nanoscale region. In particular,  
the interference between FGs due to Coulomb interaction is emerging as one of the largest 
obstacles for FG memories~\cite{Ichige,Lee,Jung}.
Stored charges in neighboring FGs interfere with one another, 
resulting in undesirable threshold voltage shifts in memory operations.
In order to reduce this interference, complicated programming sequences are 
carried out in the current commercial FG arrays.

\begin{figure}[t]
%h=here, t=top, b=bottom, p=separate figure page
\begin{center}%\leavevmode
\includegraphics[width=8.5cm]{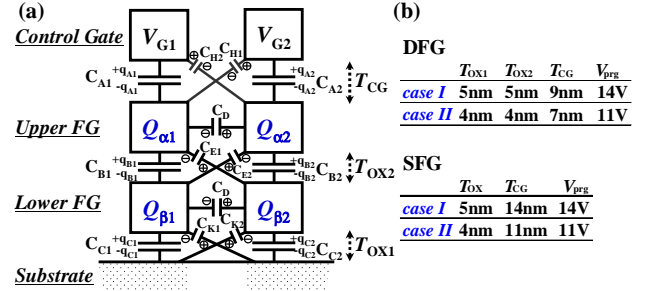}
\vspace*{3.5cm}
\caption{(a) Double floating-gate (DFG) structure. $T_{\rm ox1}$, $T_{\rm ox2}$ and $T_{\rm CG}$ are 
oxide thickness between lower FG and substrate, between two FGs, and between 
upper FG and control gate, respectively.  $q_{Ai}$, $q_{Bi}$, {\it etc.} are stored charges
in those capacitances. (b) We consider two cases of device parameters. } 
\label{DFG}
\end{center}
\end{figure}

In the case of locally charged materials such as dielectric materials, it is evident that the electric dipoles 
exist stably within mutual strong Coulomb interaction. This leads us to consider whether we can construct 
an ``artificial dipole'' using a FG system. One of the 
candidates might be a stacked double floating-gate (DFG) structure ~\cite{jjap}, in which 
an additional FG exists between the FG and the control gate in the conventional FG array 
as shown in Fig.~\ref{DFG}(a).
Moreover, if the FG becomes as small as a quantum dot~\cite{Single,Fujisawa}, 
DFG can be used as a qubit~\cite{tana}, which is a basic element of a quantum computer~\cite{DiVincenzo}.
Therefore, it is important to clarify the fundamental properties of the DFG structure. 
The purpose of this paper is to numerically compare the retention time of the DFG structure 
with that of the conventional single FG (SFG) structure in Si/SiO${}_2$ system, 
including an interference effect between two cells.
We clarify the unique transient behavior of DFG owing to the existence of the additional FG.  
In order to compare DFG with SFG impartially, we adopt
the same equivalent oxide thickness~(EOT) for both structures. 
We also compare read disturbs of both FG structures.
Finally, we derive an analytical form of charging energy of DFG and SFG as a function 
of gate voltages and electron charges.

%===================================%
%\section
{\it Formulation.---}We calculate transient behaviors
of a cell where length $L$ and width $W$ of each FG and a distance 
between neighboring cells $X_{\rm D}$ are set equal as $L=W=X_{\rm D}=23$ nm 
and the height of all FGs is $Z=50$ nm for two cases of oxide thickness (Fig.~\ref{DFG}(b)). 
(We obtain similar results for the $L=W=$11nm case.) 
We take dielectric constants and effective mass of Si and oxide SiO${}_2$ as 
$\epsilon_{\rm Si}=11.7$, $\epsilon_{\rm ox}=3.9$, and 
$m_{\rm Si}=0.19$, $m_{\rm ox}=0.5$, respectively.
The barrier height of SiO${}_2$ is $\Phi_b=2.9$eV. 
The capacitances are defined by 
$C_{\rm A}=\epsilon_{\rm Si} LW /(T_{\rm CG}+\gamma Z/2)$, 
$C_{\rm B}=\epsilon_{\rm Si} LW /(T_{\rm ox2}+\gamma Z )$, 
$C_{\rm C}=\epsilon_{\rm Si} LW /(T_{\rm ox1}+\gamma Z/2)$,
$C_{\rm D}=\epsilon_{\rm Si} ZW /(X_{\rm D}+\gamma L)$, 
$C_{\rm E}=\epsilon_{\rm Si} ZW /X_{\rm E}$,
$C_{\rm H}=\epsilon_{\rm Si} ZW /X_{\rm H}$ and 
$C_{\rm K}=\epsilon_{\rm Si} ZW /X_{\rm K}$ 
with % $\gamma\equiv\epsilon_{\rm ox}/\epsilon_{\rm Si}\approx 0.3 $ and 
$X_{\rm E}=\sqrt{(X_{\rm D}+\gamma L)^2+(T_{\rm ox2}+\gamma Z)^2}$,
$X_{\rm H}=\sqrt{(X_{\rm D}+\gamma L)^2+(T_{\rm CG}+\gamma Z/2)^2}$ and
$X_{\rm K}=\sqrt{(X_{\rm D}+\gamma L)^2+(T_{\rm CG}+\gamma Z/2)^2}$ 
where 
$T_{\rm ox1}$, $T_{\rm ox2}$ and $T_{\rm CG}$ are 
oxide thickness between lower FG and substrate, between two FGs, and between 
upper FG and control gate, respectively~(Fig.~1). 
 $\gamma=\epsilon_{\rm ox}/\epsilon_{\rm Si}$ is a penetration 
effect of small FGs~\cite{Tiwari}.
Because we use SiO${}_2$ for all tunneling barriers,  equal EOT 
means that the total thickness of barriers is the same, namely, 
$T_{\rm ox1}^{\rm DFG}+T_{\rm ox2}^{\rm DFG}+T_{\rm CG}^{\rm DFG}=T_{\rm ox}^{\rm SFG}+T_{\rm CG}^{\rm SFG}$.

%\vspace*{0.66cm}
%%%%%%%%%%%%% Fig.1

Transient calculation is carried out as follows~\cite{Pavan}. (i) For given initial charges $Q_{\alpha_i}$, $Q_{\beta_i} $ 
($i=1$ and $i=2$ indicate left cell and right cell. $\alpha$ and 
$\beta$ indicate upper FG and lower FG, respectively), potential energies of FGs
$V_{\alpha_i}$, $V_{\beta_i}$ ($i=1,2)$ are obtained by solving the matrix equations:~\cite{Pavan},
\begin{eqnarray}
Q_{\alpha_i}  &= & C_{{\rm A}_i} (V_{\alpha_i} -V_{{\rm G}_i})+C_{{\rm B}_i} (V_{\alpha_i} -V_{\beta_i})
+C_{\rm D} (V_{\alpha_i} -V_{\alpha_{\bar{i}}}) 
\nonumber \\
&+& C_{{\rm E}_i} (V_{\alpha_i} -V_{\beta_{\bar{i}}})+C_{{\rm H}_i} (V_{\alpha_i} -V_{G_{\bar{i}}}) 
\nonumber \\
Q_{\beta_i}  &= & C_{{\rm B}_i} (V_{\beta_i} -V_{\alpha_i})+C_{{\rm C}_i} (V_{\beta_i} -V_{\rm sub})
+C_{{\rm E}_i} (V_{\beta_i} -V_{\alpha_{\bar{i}}}) 
\nonumber \\
&+&C_{\rm D} (V_{\beta_i} -V_{\beta_{\bar{i}}})+C_{{\rm K}_i} (V_{\beta_i} -V_{\rm sub}).
\label{Voltage}
\end{eqnarray}
($\bar{i}=2,1$ when $i=1,2$). $V_{\rm sub}$ is a substrate bias and we set $V_{\rm sub}=0$.
(ii) Once potential energies of FGs are determined, electric field applied on 
each oxide is calculated as the difference of potential energies of FGs. For example, 
electric field between the stacked FGs is given by $E_{1i}=(V_{\beta_i}-V_{\rm sub})/T_{\rm ox1}$. 
(iii) Current through each oxide is calculated by a direct tunneling model from applied electric field $E$ 
as  
$ %
J(E)\!=\!AE^2 \exp \{-{B[ 1-(1- ET_{\rm ox}/\Phi_b )^{3/2}]}/{E}\}
% J(E)\!=\!AE^2 \exp \left\{-{B\left[ 1-\left(1- ET_{\rm ox}/\Phi_b \right)^{3/2}\right]}/{E}\right\}
$ %
with $A=e^3 m_{\rm Si} /(16 \pi^2\hbar m_{\rm ox}\Phi_b)$ and 
$B=4 \sqrt{2m_{\rm ox}}\Phi_b^{3/2}/(3\hbar e)$~\cite{Schuegraf}.
(iv) Then,  new charge distribution is obtained, namely, $Q_{\beta_i}\Rightarrow Q_{\beta_i}+(J_{1i}-J_{2i})dt$ 
with the current that flows through the lowest tunneling oxide ($J_{1i}$) and the middle tunneling oxide ($J_{2i}$) during time $dt$.
We repeat this calculation until charge distribution is stabilized by adjusting small time advance $dt$.
To determine stored charge and WRITE/ERASE process for a given gate voltage $V_{\rm prg}$, we start from 
trial charge (10${}^{-6}$~C/cm${}^2$) and repeatedly apply $V_{\rm prg}$ and $-V_{\rm prg}$ a couple of times.
The retention behavior is described under $V_{\rm G}=0$, starting from the stored charges.

%===============================================%
%===============================================%
%\section
{\it Numerical results.---}
First, we found that DFG 
is more stable when $T_{\rm ox1}=T_{\rm ox2}$. 
For example, charge distribution 
of $T_{\rm ox1}=T_{\rm ox2}=5$~nm DFG begins to change later than 
that of $T_{\rm ox1}=6$~nm and $T_{\rm ox2}=4$~nm DFG. Thus, we consider DFG with $T_{\rm ox1}=T_{\rm ox2}$.
This is because charge distribution begins to change through the thinnest tunneling oxide.

%%%%%%%%%%%%% Fig.2
\begin{figure}[t]
%h=here, t=top, b=bottom, p=separate figure page
\begin{center}%\leavevmod
\includegraphics[width=8cm]{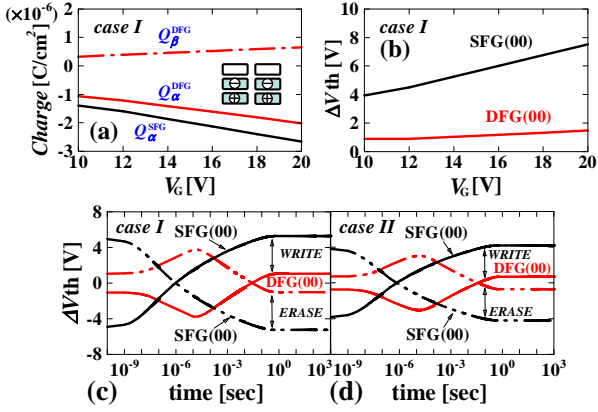}
\vspace*{5.2cm}
\caption{
(a) Stored charge in ``00" state (schematic of charge distribution of DFGs shown in inset) and 
(b) threshold voltage shift $\Delta V_{\rm th}$ as a function of gate voltage for case I.
(c) Program and erase characteristics of case I and (d) those of case II.}
\label{charge}
\end{center}
\end{figure}

%%%%%%%%%%%%% Fig.3
\begin{figure}%[t]
%h=here, t=top, b=bottom, p=separate figure page
\begin{center}%\leavevmod
\includegraphics[width=8cm]{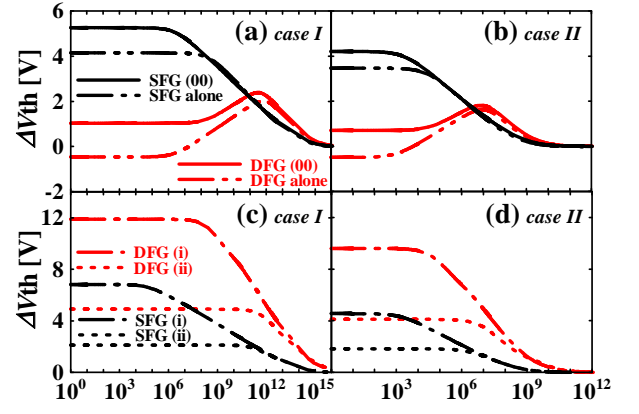}
\vspace*{5.0cm}
\caption{Transient behavior of $\Delta V_{\rm th}$ 
(retention characteristics). (a)(b) $\Delta V_{\rm th}$ of (00) and single cell (``alone") states for case I ((a)) and case II ((b)).
(c)(d) $\Delta V_{\rm th}$ of the left cell for (01) states. 
(i) $V_{G_2}=0$; $V_{G_1}=-V_{\rm prg}$ for DFG and $V_{G_1}=V_{\rm prg}$ for SFG.
(ii) $V_{G_2}=-V_{\rm prg}/2$; $V_{G_1}=-V_{\rm prg}$ for both DFG and SFG.
} 
\label{retention}
\end{center}
\end{figure}

Figure~\ref{charge} (a) shows stored charges for  programmed ``00" states in case I. 
(We obtain a similar behavior for case II.) 
We define ``0" state as a negative charge stored state (programmed state) such as 
$Q_\alpha+Q_\beta <0$ and ``1" as a charge unstored state such as $Q_\alpha+Q_\beta >0$. 
For DFG, the upper FG stores negative charges similar to the FG of  SFG, 
whereas the lower FG stores positive charges as if DFG constructs an ``artificial electric dipole". 
Figure~\ref{charge} (b) shows the threshold voltage shift $\Delta V_{\rm th}$ for  DFG and SFG.
For given charges $Q_\alpha$ and $Q_\beta$, $\Delta V_{\rm th}$ is obtained by Eqs.(\ref{Voltage}) 
as a gate voltage shift when $E_{1,i}=0$. 
For ``00" state, $\Delta V_{\rm th}$ is given when $E_{1,1}=E_{1,2}=0$ and, 
for ``01" state, $\Delta V_{\rm th}$ is given only for the left cell when $E_{1,1}=0=V_{\rm G2}$.
We can see that the magnitude of  $\Delta V_{\rm th}$
of the DFG is one-fourth smaller than that of the SFG. 
This is because positive charge and negative charge cancel electric fields with each other and
the electric field outside the dipole structure of DFG is weaker than that 
outside a single charge structure of SFG.  
This indicates that DFG is less appropriate for memories using multi-levels than SFG.
% Figures~\ref{charge}(c)(d) show WRITE/ERASE speed behavior for both DFG and SFG. 
% We can see that the WRITE/ERASE speeds are almost the same in both structures. 
In Figs.~\ref{charge}(c)(d), we show that the WRITE/ERASE speeds are almost the same in both structures. 
This is because the speed is mainly determined by the same $T_{\rm ox1}$. 
The peaks of DFG in the figures originate from 
different changes of $Q_\alpha$ and $Q_\beta$ and 
appear when the sign of $Q_\alpha+Q_\beta$ is changed.
Because the charge distribution in our model is symmetric for $V_{\rm G}>0$ and 
$V_{\rm G}<0$, the smaller memory window of DFG in Figs.~\ref{charge}(c)(d) 
corresponds to twice the $\Delta V_{\rm th}$ in Fig.~\ref{charge}(b).

Figures~\ref{retention} (a)(b) show retention characteristics, that is,  
transient degradations of $\Delta V_{\rm th}$
of DFG and SFG for ``00" and ``alone" states.
We can see that  $\Delta V_{\rm th}$ of DFG exceeds that of SFG at $\sim 10^{10}$~sec for case I
and at  $\sim 10^{6}$~sec for case II. Thus,
retention time of DFG in ``00" state is longer than that of  SFG.
The peak of DFG $\Delta V_{\rm th}$ appears when $Q_\alpha+Q_\beta$ begins to decrease. 
Note that $\Delta V_{\rm th}$ of ``alone" state starts from negative region. 
Thus, the interference between cells is effective for DFG.
Programming voltage for (01) state is different from that for (00) state. In Fig.~\ref{retention} (c)(d),
negative voltage is applied to store negative charge for DFG and SFG of $V_{{\rm G}_2}\neq 0$.
These are results of the complicated electromagnetic fields produced by Coulomb interactions among FGs.
In any case, $\Delta V_{\rm th}$s of ``01" state in DFG become larger than those of SFG 
even in this thin $T_{\rm ox1}$ ($ \le 5$nm) region. (We have similar relation between DFG and SFG 
for $T_{\rm ox1}^{\rm DFG}=T_{\rm ox2}^{\rm DFG}=T_{\rm ox}^{\rm SFG}=7$nm, $T_{\rm CG}^{\rm DFG}=13$ and 
$T_{\rm CG}^{\rm SFG}=20$nm~\cite{EPAPS}.) % (figures not shown).)
Once dipole is formed in DFG, its surrounding electric field is weaker than that of SFG. Thus, 
DFG weakly couples with its environment, resulting in longer retention time. 
The interference between two cells is considered to stabilize the charge redistribution of horizontal directions.

\begin{figure}[t]
%h=here, t=top, b=bottom, p=separate figure page
\begin{center}%\leavevmod
\includegraphics[width=5.5cm]{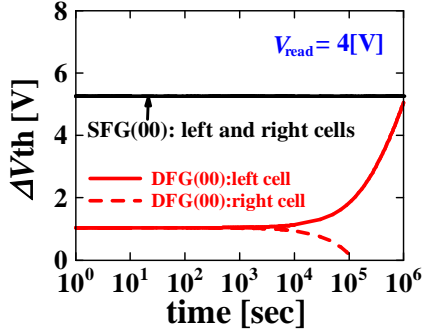}
\vspace*{4.0cm}
\caption{Read disturb at $V_{\rm read}=4$~V starting from ``00" states for case I. 
$V_{\rm read}$ is applied to the left cell. 
Note that $\Delta V_{\rm th}$s for SFG cells do not change during this time scale.} %[091207] 
\label{read}
\end{center}
\end{figure}

{\it Read disturb.---}
Figure~\ref{read} shows retention characteristics in which gate bias $V_{\rm G_1}=V_{\rm read}=4$~V  
is applied on the left cell while $V_{\rm G_2}=0$, assuming that the left cell is read by $V_{\rm read}$ (read disturb process). 
We obtain a similar behavior for case II.
As can be seen, DFG changes earlier than SFG, but because 
the reading process is carried out in millisecond order, this 
weakness does not affect the practical usage of DFG.
 
%%%%%%%%%%%%% Fig.5

%=====================================%
%\section
{\it Charging energy.---}
In the near future of the down-scaling of FGs, the number of electrons in the FGs 
is reduced and countable, resulting in the region of single-electronics~\cite{Fujiwara}.
In this region, an analytical form of charging energy as functions of electrical charges and gate biases 
is required to describe an experimental stability diagram of 
electron charge distribution as shown in Ref.~\cite{Leo}.
Here we provide an analytical form of the charging energy of two cells (Fig.~\ref{DFG}) by using 
a capacitance network model.

The charging energy of the system can be expressed 
by summing charging energy of all capacitors such as $U_{\rm ch}=\sum_{l} q_l^2/(2C_l)
-\sum_{l'} q_{l'} V_{l'}$ where $q_l$ shows charge of each capacitor and $q_{l'}$ shows 
charge of capacitor that connects to gate voltage~(see Fig.~1). 
By using Lagrange multipliers similar to Ref.~\cite{tana}, we have the charging energy 
of two DFGs as 
$U_{\rm DFG}=U_{\rm I}+U_{\rm II}-\sum_{i=1}^2 \{ C_{{\rm A}_i} V_{{\rm G}_i}^2+ C_{{\rm H}_i} V_{{\rm G}_{\bar{i}}}^2 \}/2$,
where
\begin{eqnarray}
U_{\rm I}&=&\frac{D_1 w_{g_2}^2+D_2 w_{g_1}^2+2C_y w_{g_1}w_{g_2}}{2[D_1D_2-C_y^2]} 
% U_{\rm I}&=&\frac{D_1 w_{g_2}^2+D_2 w_{g_1}^2+2C_y w_{g_1}w_{g_2}}{2[D_1D_2-C_y^2]\Delta^2} 
\nonumber \\
U_{\rm II}&=&(C_{a_2} v_{g_1}^2 +C_{a_1} v_{g_2}^2+2C_{\rm D}  v_{g_1}v_{g_2} )/(2{\Delta}) 
\label{DFGcharge}
\end{eqnarray}
with
$v_{g_i} =C_{{\rm A}_i}V_{{\rm G}_i} +C_{{\rm H}_{\bar{i}}}V_{{\rm G}_{\bar{i}}}-Q_{\alpha_i}^{\rm DFG}$ and  
\begin{eqnarray}
w_{g_i}&=&\{[C_{\rm D}C_{{\rm E}_{\bar{i}}} + C_{a_{\bar{i}}}C_{{\rm B}_i}]/\Delta \}v_{g_i} \nonumber \\
&+&\{[C_{\rm D}C_{{\rm B}_i} +  C_{a_i}C_{{\rm E}_{\bar{i}}}]/\Delta \} v_{g_{\bar{i}}} -   Q_{\beta_i}^{\rm DFG}, 
\end{eqnarray} %\nonumber \\
including $C_{a_i} = C_{{\rm A}_i}+C_{{\rm B}_i}+C_{\rm D}+C_{{\rm E}_i}+C_{{\rm H}_{\bar{i}}}$,   
%% $C_{a_2}=C_{A_2}+C_{B_2}+C_D+C_{E_2}+C_H$ and 
$C_{b_i}=C_{{\rm B}_i}+C_{{\rm C}_i}+C_{\rm G}+C_{{\rm E}_{\bar{i}}}+C_{{\rm K}_{\bar{i}}}$, 
%% $C_{b_2}=C_{B_2}+C_{C_2}+C_G+C_E+C_K$,
$\Delta\equiv C_{a_1}C_{a_2}-C_{\rm D}^2$,  
$D_i\equiv C_{b_i}-[C_{a_{\bar{i}}}C_{{\rm B}_i}^2+C_{a_i}C_{{\rm E}_{\bar{i}}}^2+2C_{\rm D} C_{{\rm E}_{\bar{i}}}C_{{\rm B}_i}]/\Delta$, 
%%--- $D_2\equiv C_{b_2}-[C_{a_1}C_{B_2}^2+C_{a_2}C_{E}^2+2C_DC_EC_{B_2}]/\Delta$, 
and $C_y\equiv C_{\rm G}+[C_{a_1}C_{{\rm B}_2}C_{\rm E_2}+C_{a_2}C_{\rm B_1}C_{\rm E_1}
+C_{\rm B_1}C_{\rm B_2}C_{\rm D}+C_{{\rm E}_1}C_{\rm E_2}C_{\rm D}]/\Delta$.
%% $w_2=[C_DC_E+C_{a_1}C_{B_2}]v_{g2}+[C_DC_{B_2} +C_{a_2}C_E]v_{g1}-\Delta Q_{\alpha_2}$,
%% $v_{g_2}=C_{A_2}V_{G_2}+C_{H}V_{G_1}-Q_{\alpha_2}$.
Note that $U_{\rm SFG}$ has the same form as $U_{\rm II}$ with capacitances replaced
by those of SFG such as $C_{{\rm A}i}^{\rm \small S}$, $C_{{\rm C}i}^{\rm \small S}$ {\it etc}.
For ``00" state ($w_g\equiv w_{g_1}=w_{g_2}$ and $v_g\equiv v_{g_1}=v_{g_2}$), we have,
\begin{equation}
U_{\rm I}= w_g^2/(D-C_y), \  U_{\rm II}=v_g^2/(C_a-C_{\rm D}) 
\label{00state}
\end{equation}
For SFG, we have $U_{\rm SFG}=v_g^2/(C_a^{\rm S}-C_{\rm D}^{\rm S})-(C_{\rm A}+C_{\rm H})V_{\rm G}^2$ where 
$C_a^{\rm S}=C_{\rm A}^{\rm S}+C_{\rm C}^{\rm S}+C_{\rm D}^{\rm S}+C_{\rm H}^{\rm S}+C_{\rm J}^{\rm S}$ 
and $v_g=(C_{\rm A}^{\rm S}+C_{\rm H}^{\rm S})V_{\rm G}-Q_{\alpha}^{\rm SFG}$.

In summary, 
we numerically showed that interfering DFGs have a longer retention time 
than SFGs, although DFGs have the disadvantage of 
smaller threshold voltage shift and smaller memory window. %[091207]
Owing to the existence of additional FG, DFG shows unique transient characteristics, 
forming an artificial electric dipole.

The authors thank A. Nishiyama, J. Koga, S. Fujita, N. Yasuda and A. Kinoshita for useful discussions.

%%%%%%%%%%%%%%%%%%%%%%%%%%%%%%%%%%%%%%%%%%%%%%

\end{document}